\documentclass[aps,prd,onecolumn,noshowpacs,preprintnumbers,eqsecnum,nofootinbib]{revtex4} 
\usepackage[reqno]{amsmath}
\usepackage{epsfig}
\usepackage{array}
\usepackage{float}
\usepackage{lscape,graphicx}
\usepackage{amssymb}
\usepackage[usenames,dvipsnames]{color}
\usepackage{ulem}
\usepackage{enumerate}
\usepackage{natbib}

\usepackage{epsfig}
\def\gs{\mathrel{
   \rlap{\raise 0.511ex \hbox{$>$}}{\lower 0.511ex \hbox{$\sim$}}}}
\def\ls{\mathrel{
   \rlap{\raise 0.511ex \hbox{$<$}}{\lower 0.511ex \hbox{$\sim$}}}}
\def\gtap{\mathrel{
   \rlap{\raise 0.511ex \hbox{$>$}}{\lower 0.511ex \hbox{$\sim$}}}}
\def\ltap{\mathrel{
   \rlap{\raise 0.511ex \hbox{$<$}}{\lower 0.511ex \hbox{$\sim$}}}}

\begin{document}

\preprint{{\it SISSA 55/2016/FISI}}
\preprint{{\it IPMU16--0167}}

\title{On the IceCube Result on $\bar{\nu}_{\mu} \to \bar{\nu}_{s}$ 
Oscillations}
\author{S.T. Petcov$^{1,2}$}
\thanks{Also at: Institute of Nuclear Research and Nuclear Energy,
Bulgarian Academy of Sciences, 1784 Sofia, Bulgaria}
\affiliation{
$^{\it 1}$SISSA/INFN, 
34136 Trieste, Italy\\
$^{\it 2}$Kavli IPMU (WPI), University of Tokyo, 277-8583
Kashiwa, Japan\\
}

\vspace{.5cm}
\hfuzz=25pt
\begin{abstract}
 We elucidate the mechanism of enhancement of the
$\bar{\nu}_{\mu} \to \bar{\nu}_{s}$ transitions at small mixing
angles of Earth-core-crossing neutrinos, which is at the basis of
the  recent result of the IceCube experiment on atmospheric muon
anti-neutrino oscillations into sterile anti-neutrino.
\end{abstract}

\maketitle
%

\vspace{-0.8cm}
\section*{}

\vspace{-0.8cm} 

Recently the IceCube collaboration 
published negative results of their search for 
oscillations of atmospheric muon anti-neutrinos  into sterile
anti-neutrinos~\cite{TheIceCube:2016oqi},
$\bar{\nu}_{\mu} \to \bar{\nu}_{s}$.
The data on $\bar{\nu}_{\mu} \to \bar{\nu}_{s}$ oscillations were
 analysed in \cite{TheIceCube:2016oqi}
in terms of the
two-neutrino mixing hypothesis, 
i.e., in terms of two parameters:
one neutrino mass squared difference 
$\Delta m^2$ and one $\nu_\mu-\nu_s$ mixing angle $\theta$ (denoted
$\theta_{24}$ in \cite{TheIceCube:2016oqi}).
The IceCube collaboration 
did not find evidence for  $\bar{\nu}_{\mu} \to \bar{\nu}_{s}$ oscillations 
and in \cite{TheIceCube:2016oqi}
exclusion limits in the 
$\Delta m^2 - \sin^22\theta$ plane were reported. 
The IceCube limits on $\sin^22\theta$ are particularly strong in the 
interval $\Delta m^2\sim (0.1 - 1.0)~{\rm eV^2}$, 
where at 99\% C.L. the upper bound on  $\sin^22\theta$ decreases 
from approximately 0.10 down to 0.04; at 90\% C.L. 
it decreases down to 0.02.   
These limits on $\sin^22\theta$ are stronger than those 
obtained in other experiments in the same  $\Delta m^2$ 
region (see, e.g., Fig. 5 in \cite{TheIceCube:2016oqi}).
In connection with this result it is stated in the Abstract of 
ref. \cite{TheIceCube:2016oqi}
that ``New exclusion limits are placed on the parameter space
... in which muon antineutrinos would experience a strong
Mikheyev-Smirnov-Wolfenstein resonant oscillations.'' 
\begin{figure}[tmb]
\begin{center}
\includegraphics[width=8.5cm,height=8.5cm]{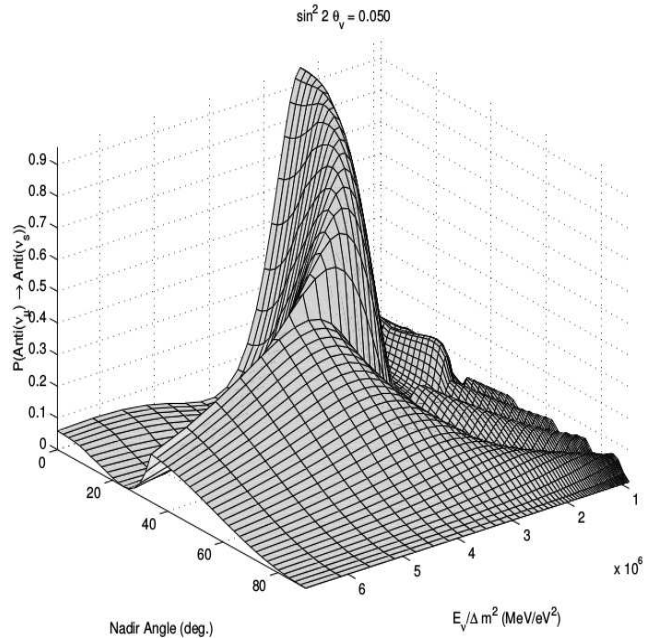}
\caption{Oscillogram of the Earth for $\bar{\nu}_{\mu} \to
\bar{\nu}_{s}$ transitions: the probability $P(\bar{\nu}_{\mu} \to
\bar{\nu}_{s})$ 
as a function of the nadir angle $h$~[degree] and 
$E/\Delta m^2~{\rm [10^6~MeV/eV^2]}$ 
for $\sin^22\theta =0.05$.
Neutrinos cross the Earth core for $0^\circ \leq h < 33.17^\circ$.
The absolute maximum of $P(\bar{\nu}_\mu \to \bar{\nu}_s) \approx 0.95$
seen in the figure at $h = 0^\circ$ is due to the Earth mantle-core
enhancement mechanism. As was shown in \cite{PRD1999}, for $h=0^\circ$
the same mechanism leads to $P(\bar{\nu}_\mu \to \bar{\nu}_s) = 1$ at
$\sin^22\theta = 0.07$ (see Table 3 in \cite{PRD1999}). The figure is
from \cite{Chizhov:1998ug}.
See \cite{Chizhov:1998ug,PRD1999} for further details. }
\label{Fig1}
\end{center}
\end{figure}
%
  
 We point out in the present article 
that the strong exclusion limit of the IceCube experiment 
on $\bar{\nu}_{\mu} \to \bar{\nu}_{s}$ oscillations
in the region of $\Delta m^2\sim (0.1 - 1.0)~{\rm eV^2}$, 
apart from the experimental 
characteristics of the IceCube detector, 
is a consequence exclusively 
of the ``Neutrino Oscillation Length Resonance-like'' (NOLR), or 
``Earth mantle-core'', enhancement of the neutrino transitions 
at small mixing angles for neutrinos crossing the Earth core 
\cite{Petcov:1998su,Chizhov:1998ug,PRL1999,PRD1999}.
This mechanism of enhancement of neutrino oscillations
differs \cite{Petcov:1998su} from the MSW one
\cite{LW78,MS85}.
The Earth mantle-core or NOLR enhancement
was found to be operative and remarkably strong 
in the $\bar{\nu}_{\mu} \to \bar{\nu}_{s}$ transitions 
at small mixing angles first in \cite{Chizhov:1998ug}
(see Fig.~\ref{Fig1}).
A complete description of the 
mechanism of Earth mantle-core enhancement of
the $\bar{\nu}_{\mu} \to \bar{\nu}_{s}$ 
and all other transitions of the 
Earth-core-crossing neutrinos subject to the 
enhancement (see
\cite{Petcov:1998su,Chizhov:1998ug}),
was presented in \cite{PRL1999,PRD1999}.
It was proven in \cite{PRL1999,PRD1999} that, 
as was suggested in \cite{Petcov:1998su}
(in connection with the discovered 
NOLR amplification 
of the $\nu_{\mu(e)} \rightarrow \nu_{e(\mu)}$ 
and $\nu_{2} \rightarrow \nu_{e}$ transitions  
at small mixing angles),  
the enhancement is caused by maximal constructive interference 
between the neutrino transition amplitudes in the Earth mantle and
in the Earth core (hence ``Earth mantle-core'' 
effect). 
Although it has resonance-like ``appearance'' in the space of
relevant parameters (see, e.g., Fig. \ref{Fig1}), 
it has no true resonance origin 
\footnote{Thus, the term 
``Neutrino Oscillation Length Resonance'' (NOLR)
used, e.g., in \cite{Petcov:1998su,Chizhov:1998ug} 
for the enhancement of interest is somewhat inaccurate.
A more precise term  is, e.g.,  
``Neutrino Oscillation Length Resonance-like'' enhancement, 
which we are using in the present article keeping the same
abbreviation - NOLR - for it.}. 
The Earth mantle-core or NOLR enhancement takes
place  at small mixing angles also in 
the $\nu_{\mu(e)} \rightarrow \nu_{e(\mu)}$, 
$\nu_{2} \to \nu_{e}$ \cite{Petcov:1998su},
and $\nu_{e} \to \nu_{s}$ \cite{Chizhov:1998ug}
transitions (see also \cite{PRL1999,PRD1999}).
For all quoted transitions, $\bar{\nu}_{\mu} \to \bar{\nu}_{s}$, 
$\nu_{\mu(e)} \rightarrow \nu_{e(\mu)}$, $\nu_{2} \to \nu_{e}$, and 
$\nu_{e} \to \nu_{s}$, it can be maximal \cite{PRL1999,PRD1999}.
As was shown in \cite{Petcov:1998su,Chizhov:1998ug,PRL1999,PRD1999},
for neutrinos crossing the Earth core 
and the listed transitions, the Earth mantle-core 
(or NOLR)  enhancement at small mixing 
angles  is significantly larger 
than the enhancement 
due to the MSW resonance \cite{LW78,MS85}
taking place in the Earth mantle or in the Earth core,  
and represents the dominant 
amplification mechanism 
for the indicated 
transitions. 

In order to explain the mechanism of the 
Earth mantle-core enhancement, it is sufficient to 
consider the case of 2-neutrino mixing
employed in the analyses performed
in \cite{TheIceCube:2016oqi}
and in \cite{Petcov:1998su,Chizhov:1998ug,PRL1999,PRD1999}, and use 
the approximation of constant densities
and electron fraction numbers of the Earth mantle, $\rho_m$ and 
$Y^m_e$, and of the Earth core, $\rho_c$ and $Y^c_e$ 
(for details see \cite{Petcov:1998su,Chizhov:1998ug,PRD1999}).
We recall that according to the 
Earth models \cite{Stacey,PREM}, 
the Earth density distribution is 
spherically symmetric and consists of 
two major density structures -
the core and the mantle - and
a certain number of substructures (shells or layers).
The Earth radius is \cite{Stacey,PREM} $R_{\oplus} = 6371$ km.
According to the Stacey model \cite{Stacey}
(PREM model \cite{PREM}), the Earth core
has a radius of $R_c= 3485.7$ (3480) km, so
the Earth mantle depth is 2885.3 (2891) km.
For a spherically symmetric Earth density
distribution, the neutrino trajectory
in the Earth is specified by the value
of the nadir angle $h$ of the trajectory.
For nadir angles $h < 33.17^\circ$
(or path lengths $L > 10665.7$ km)
neutrinos cross the Earth core.
According to the Earth models \cite{Stacey,PREM},
the mean density  of the core is
larger approximately by a factor of 2.5                                       
than the mean density in the mantle and the change of
the density from the mantle to the core can well be
approximated by a step function.
The mean densities and electron fraction numbers
of the Earth mantle and core 
in the Stacey Earth model read ~\cite{Stacey}:
$\rho_{man}\approx 4.5~{\rm g}/{\rm cm^3}$,
$Y^{man}_e\approx 0.49$, and $\rho_c\approx 11.5~{\rm g}/{\rm cm^3}$,
$Y^c_e\approx 0.467$.
The values of $\rho_{man}$, $Y^{man}_e$, $\rho_c$,
$Y^c_e$, $R_c$ and $R_m$ in the PREM Earth model
\cite{PREM} are very similar
%
%
and lead to the same results.
The  density and the electron fraction number
change relatively little around
the indicated mean values
along the trajectories of neutrinos
which cross a substantial part of the
Earth mantle, or the mantle and the core,
and the two-layer constant density approximation
was shown to be sufficiently accurate in what
concerns the calculation of neutrino oscillation
probabilities 
\cite{PKSP3nu88,Petcov:1998su,Chizhov:1998ug}
(and references quoted in \cite{Petcov:1998su,Chizhov:1998ug})
in a large number of specific cases.
This is related to the fact
that the relatively small changes of density
along the path of the neutrinos in the mantle
or in the core take place over path lengths which
are typically considerably smaller than
the corresponding oscillation length
in matter.

 In the case of 2-neutrino mixing, relevant for our discussion, 
 we have
 $P(\bar{\nu}_\mu \to \bar{\nu}_\mu) = (1 - P(\bar{\nu}_\mu \to \bar{\nu}_s))$, 
$P(\bar{\nu}_\mu \to \bar{\nu}_\mu)$ and  $P(\bar{\nu}_\mu \to \bar{\nu}_s)$ 
being the $\bar{\nu}_{\mu}$ survival and 
the  $\bar{\nu}_{\mu} \to \bar{\nu}_{s}$ transition probabilities,
respectively.
The maximal Earth mantle-core enhancement of the $\bar{\nu}_{\mu}
\to \bar{\nu}_s$ transition probability leading to a total neutrino
conversion, 
$P(\bar{\nu}_{\mu} \to \bar{\nu}_s) = 1$, 
is realized when a specific synchronisation
between the oscillation phases (or oscillation lengths
\cite{Petcov:1998su})
 in the Earth mantle $2\phi' =  X' \Delta E'$, and in
the Earth core, $2\phi'' = X'' \Delta E''$, $X'$ ($X''$) and $\Delta
E'$ ($\Delta E''$) being the neutrino path length and
the difference between the energies of the two neutrino matter
eigenstates in any of the two mantle layers (in the core),
takes place \cite{PRL1999}:
\begin{equation}
\label{A}
\tag{1}
    \left\{\begin{array}{lcl}
            \tan\phi' & =
            & \pm\, \sqrt{\dfrac{-\cos 2\theta''_m}
                {\cos(2\theta''_m-4\theta'_m)}}~, \\
            \tan\phi'' & =
            & \pm\,\dfrac{\cos 2\theta'_m}{\sqrt{-\cos
            2\theta''_m\cos(2\theta''_m-4\theta'_m)}}~,
          \end{array}
    \right.
\end{equation}
%
where 
\footnote{The conditions of synchronisation between the
 oscillation lengths in the Earth mantle and the Earth core 
found in \cite{Petcov:1998su}, which are a particular case of the 
general conditions derived in 
\cite{PRL1999,PRD1999} and given in eq.~(\ref{A}),
are at the origin of the term ``Neutrino Oscillation Length
Resonance'' (NOLR) introduced in \cite{Petcov:1998su} 
and used, e.g., in \cite{Chizhov:1998ug}.}
$\theta'_m$ and $\theta''_m$ are the mixing angles in the
Earth mantle and core, respectively, and the signs in the
expressions for $\tan\phi'$ and $\tan\phi''$ are correlated.
Solutions of eq.~(\ref{A})
corresponding to the Earth mantle-core enhancement and
leading to $P(\bar{\nu}_{\mu} \to \bar{\nu}_s) = 1$
are possible only in the region
\cite{PRL1999,PRD1999}
\begin{equation}
\label{region}
\tag{2}
    \left\{\begin{array}{l}
             \cos 2\theta''_m \leq 0~, \\
             \cos(2\theta''_m - 4\theta'_m) \geq 0~.
           \end{array}
    \right.
\end{equation}
%

\begin{figure}[tmb]
\begin{center}
\includegraphics[width=8cm,height=8cm]{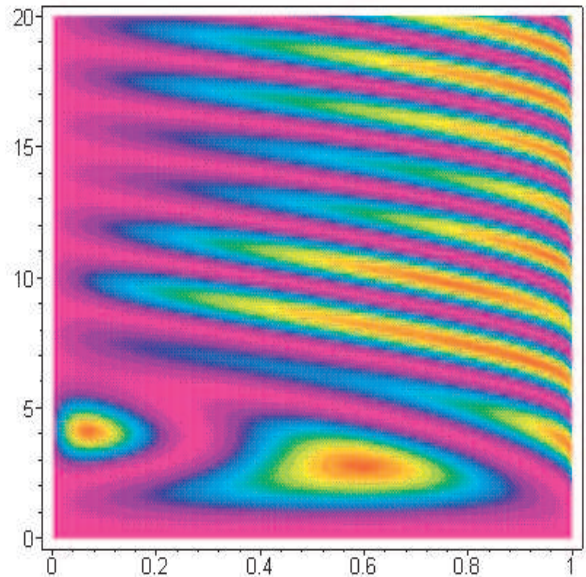}
\caption{The probability $P(\bar{\nu}_\mu \to \bar{\nu}_s)$ for
Earth-center-crossing (atmospheric) neutrinos ($h = 0^{\circ}$), as
a function of $\sin^22\theta$ (horizontal axis) and 
$\Delta m^2/E~[10^{-7}{\rm eV^2/MeV}]$ 
(vertical axis). 
The ten different colors
correspond to values of $P(\bar{\nu}_\mu \to \bar{\nu}_s)$ in the
intervals: 0.0 - 0.1 (violet); 0.1 - 0.2 (dark blue); ...; 0.9 - 1.0
(dark red). The points of total neutrino conversion (in the dark red
regions), $P(\bar{\nu}_\mu \to \bar{\nu}_s) = 1$, correspond to
solutions of eq. (\ref{A}). The figure is from
hep-ph/9903424 quoted in \cite{PRD1999}.
For further details see \cite{PRD1999}. } 
\label{Fig2}
\end{center}
\end{figure}

\begin{figure}[tmb]
\begin{center}
\begin{tabular}{cc}
\includegraphics[width=8cm,height=8cm]{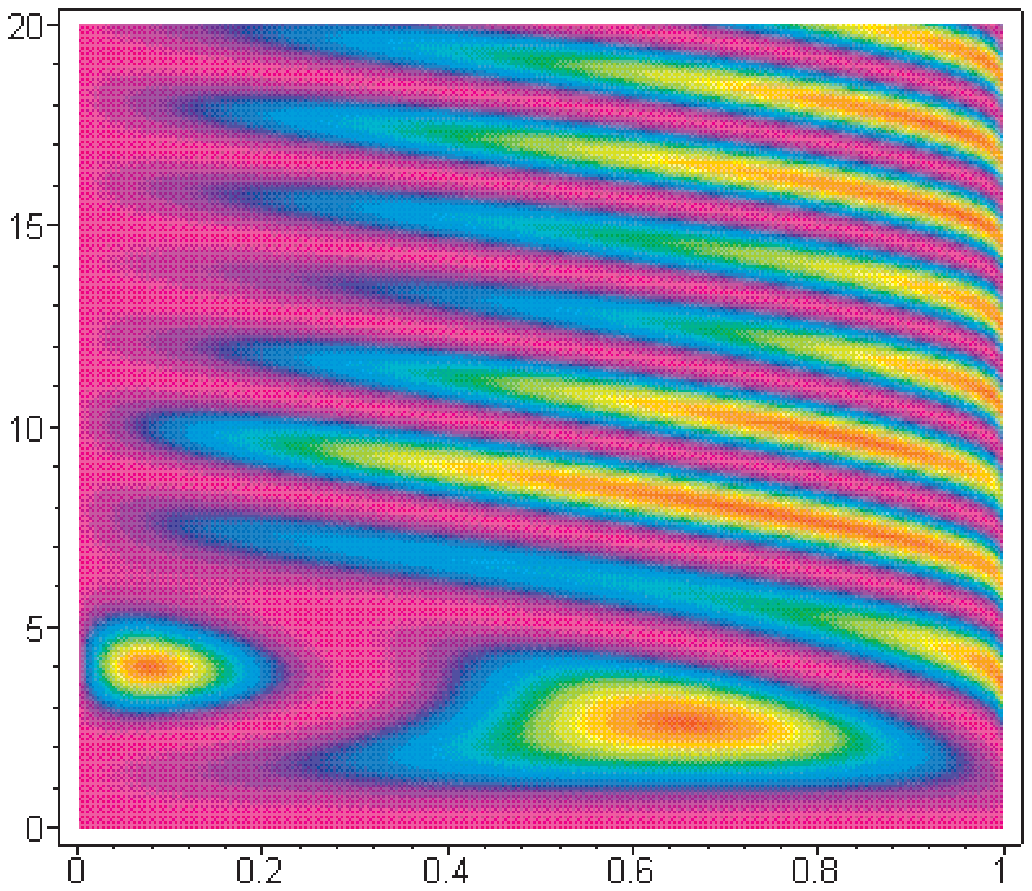}&
\includegraphics[width=8cm,height=8cm]{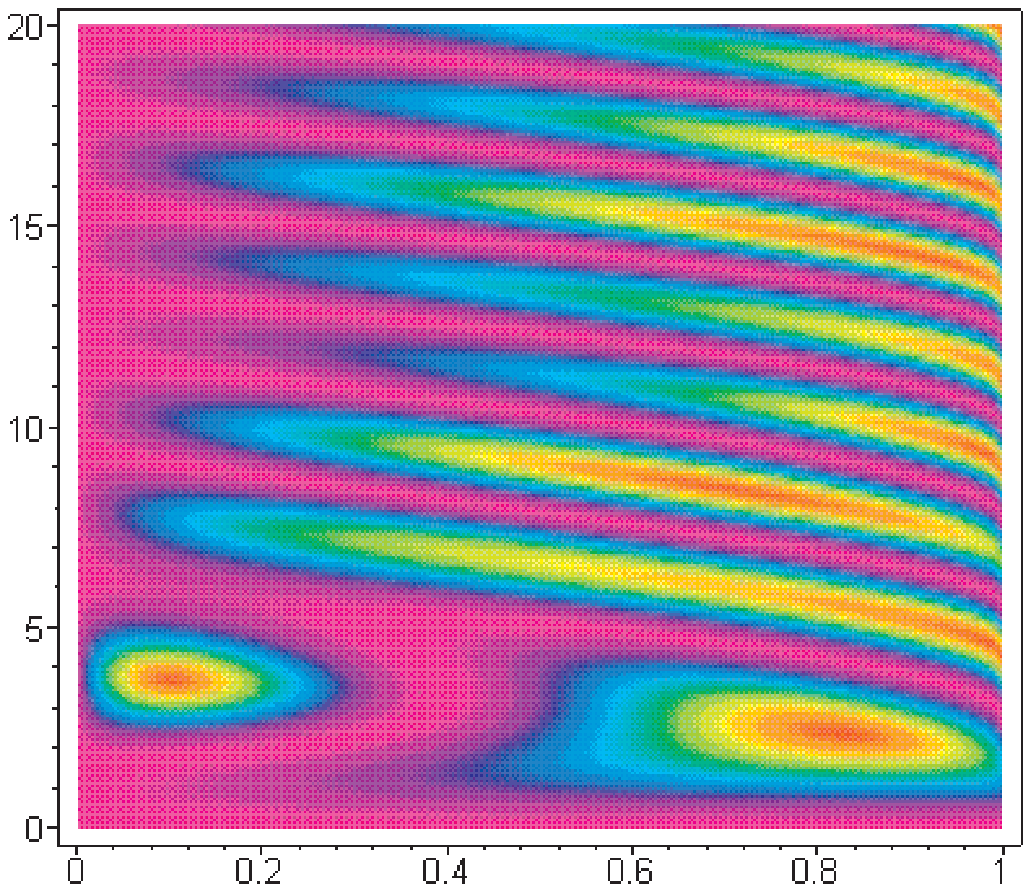}
\end{tabular}
\caption{ The same as in Fig.~\ref{Fig2} for $h = 13^{\circ}$ 
(left panel) and $h = 23^{\circ}$ (right panel). 
(From hep-ph/9903424 quoted in 
\cite{PRD1999} and \cite{PRD1999}.) } 
\label{Fig34}
\end{center}
\end{figure}

At small mixing angles of interest 
the maximal Earth mantle-core enhancement 
solutions of eq. (\ref{A}) implying 
$P(\bar{\nu}_{\mu} \to \bar{\nu}_s) = 1$
exist at all nadir angles, corresponding to 
Earth-core-crossing neutrino trajectories, 
$0^\circ \leq h < 33.17^\circ$.
The solutions of eq.~(\ref{A}) in the 
case of $\bar{\nu}_{\mu} \to \bar{\nu}_s$  oscillations for, e.g., 
$h = 0^\circ$, $13^\circ$, $23^\circ$, $30^\circ$, 
can be found in Table 3 of \cite{PRD1999};
for $h = 0^\circ$, $13^\circ$, $23^\circ$ they are 
shown in Figs. 12 - 14 in \cite{PRD1999} (hep-ph/9903424),
which are reproduced here as Figs. \ref{Fig2} - \ref{Fig34}.
The strong IceCube exclusion limits 
in the region $\sin^22\theta \ltap 0.10$, 
$\Delta m^2 \sim (0.1 - 1.0)$ eV$^2$,
which at 99\% (90\%) C.L. extend down 
to $\sin^2 2\theta \approx$ 0.04 (0.02) 
\cite{TheIceCube:2016oqi}
and are better than those 
reported by other experiments, 
are obtained due to the 
strong Earth mantle-core enhancement of 
$P(\bar{\nu}_{\mu} \to \bar{\nu}_s)$ 
discovered in 
\cite{Chizhov:1998ug} 
and related \cite{PRD1999} to the  
solutions of eq.~(\ref{A}) at small mixing angles 
and $E/\Delta m^2 \sim 3$~TeV/eV$^2$. 
Indeed, at, e.g., $h = 0^\circ$, $13^\circ$, $23^\circ$, 
$30^\circ$, the solutions of eq. (\ref{A}) 
under discussion corresponding to 
$P(\bar{\nu}_{\mu} \to \bar{\nu}_s) = 1$ 
take place at 
$\sin^22\theta = 0.070$, 0.077, 0.101, 0.150, 
respectively (Table 3
in \cite{PRD1999}). 
The maximum disappearance of 
muon antineutrinos 
at $E \approx 2.7$~TeV shown in
Fig.~1 (middle panel) in \cite{TheIceCube:2016oqi}
and found at the best-fit point
$\Delta m^2=1$ eV$^2$ and $\sin^2 2\theta=0.1$, 
for example,
exactly corresponds to the
solution in Table 3 of \cite{PRD1999} 
at $h=23^\circ$ (or $\cos\theta_z\simeq -0.92$, $\theta_z$
being the zenith angle) and 
$\Delta m^2/E = 3.70\times 10^{-7}$~eV$^2$/MeV, seen also 
in the right panel of Fig. \ref{Fig34}. 
%

%
%

  In contrast to the Earth mantle-core (or NOLR)
enhancement, the MSW enhancement of neutrino mixing in matter
has a genuine resonance character \cite{LW78,MS85}.
As can be shown, the MSW resonant effect 
plays a sub-dominant (if not insignificant) 
role for the $\bar{\nu}_{\mu} \to
\bar{\nu}_{s}$ IceCube results \cite{TheIceCube:2016oqi} 
at small mixing angles. 
Indeed, for the values of $\sin^22\theta \ltap 0.1$ 
and $\Delta m^2 \cong (0.1 - 1.0)~{\rm eV^2}$ of interest,
the  $\bar{\nu}_{\mu} \to \bar{\nu}_{s}$ transitions 
are strongly suppressed for the Earth-core-crossing 
neutrinos when the MSW resonance 
condition \cite{MS85,Barger:1980tf} 
is fulfilled in the mantle because 
for the quoted relevant values of $\sin^22\theta$ and 
$\Delta m^2$, 
i) the neutrino oscillation length in the Earth mantle is much larger 
than the distance traveled by neutrinos in the mantle, and 
ii) the neutrino mixing in the Earth core, more precisely the 
parameter $\sin^22\theta''_m$, is strongly suppressed due to the 
(neutron) density of the core being significantly larger 
than the (neutron) density in the 
mantle 
\footnote{The analytic expression for the  
$\bar{\nu}_{\mu} \to \bar{\nu}_{s}$ transition 
probability in the two layer constant density approximation 
for the Earth density distribution  
can formally be obtained from eq. (7) in \cite{Petcov:1998su} 
by setting the angle $\theta$ in eq. (7) to zero and 
by taking into account 
that the matter potential for the transitions of interest 
is given by $\sqrt{2} G_F N_n/2$, where 
$G_F$ and  $N_n$ are the Fermi constant and the 
neutron number density of matter.
The mean neutron number density of the Earth mantle and 
of the Earth core, $N_n^{man} \propto \rho_{man}(1 - Y^{man}_e)$ and 
$N_n^{c} \propto \rho_{c}(1 - Y^{c}_e)$, 
have the values \cite{Stacey}:
$N_n^{man} \cong 2.3~{\rm N_A~cm^{-3}}$ and 
$N_n^{c} \cong 6.1~{\rm N_A~cm^{-3}}$, ${\rm N_A}$ being 
the Avogadro's number.
}.
The same conclusion is valid when the MSW resonance 
takes place in the Earth core.  
The absence of significant MSW amplification  
of $P(\bar{\nu}_{\mu}\to \bar{\nu}_s)$ 
at small mixing angles for 
Earth-core-crossing neutrinos 
is evident in Figs. \ref{Fig2} 
and \ref{Fig34}, where the only 
``red spot'' present at $\sin^22\theta \ltap 0.15$ 
in each of the three figures is that associated with 
the maximal Earth mantle-core enhancement of 
 $P(\bar{\nu}_{\mu}\to \bar{\nu}_s)$ 
at $\Delta m^2/E \cong (3.7 - 4.0)\times 10^{-7}~{\rm eV^2/MeV}$.

When neutrinos traverse only the mantle, 
the MSW resonance in the mantle 
can enhance somewhat the 
probability $P(\bar{\nu}_{\mu}\to \bar{\nu}_s)$, 
with respect to its rather small vacuum value, 
but only for a limited set of neutrino trajectories 
located very close to the core-mantle boundary.
In addition the MSW enhancement is 
noticeably smaller 
than that due to the Earth mantle-core 
effect for the  core-crossing neutrinos.
The local enhancement of
the  $\bar{\nu}_{\mu} \to \bar{\nu}_{s}$ transition,
seen in Fig.~1 (middle panel) in \cite{TheIceCube:2016oqi}
at  $\cos\theta_z \geq -0.84$ (or $h \geq 33.17^\circ$) 
is due to the MSW resonance in the
mantle and illustrates this conclusion. This MSW local
minimum of $P(\bar{\nu}_{\mu} \to \bar{\nu}_\mu)$ (local maximum of  
$P(\bar{\nu}_{\mu} \to \bar{\nu}_s)$)
is reached at $E \sim 4.4$ TeV
and at it $P(\bar{\nu}_{\mu}\to \bar{\nu}_\mu)\sim 0.40$
($P(\bar{\nu}_{\mu} \to \bar{\nu}_s)\sim 0.60$). 
Therefore the statement in the Abstract of \cite{TheIceCube:2016oqi}
that ``New exclusion limits are placed on the parameter space
... in which muon antineutrinos would experience a strong
Mikheyev-Smirnov-Wolfenstein resonant oscillation.'' is at least
misleading (if not incorrect).

{\bf Acknowledgements.} 
The author would like to use this opportunity to
thank M. Maris and M.V. Chizhov for the 
collaborations respectively on the articles
\cite{Chizhov:1998ug} and \cite{Chizhov:1998ug,PRL1999,PRD1999}.
This work was supported in part by the INFN
program on Theoretical Astroparticle Physics (TASP), by the research
grant  2012CPPYP7
under the program  PRIN 2012 funded by the Italian MIUR,
by the European Union Hîrizon 2020 research and innovation programme
under the  Marie Sklodowska-Curie grants 674896 and 690575, and by
the World Premier International Research Center Initiative (WPI
Initiative), MEXT, Japan.

\end{document}